\documentstyle[preprint,aps]{revtex}
\newcommand{\be}{\begin{equation}} \newcommand{\ee}{\end{equation}} 
\newcommand{\bea}{\begin{eqnarray}}\newcommand{\eea}{\end{eqnarray}}

\begin{document}
\draft
\preprint{OCHA-SP-00-02, hep-th/0007208}
\title{ Super-Calogero-Moser-Sutherland systems and free super-oscillators
: a mapping}
\author{Pijush K. Ghosh$^{*}$}
\address{ 
Department of Physics,
Ochanomizu University,\\
2-1-1 Ohtsuka, Bunkyo-ku,
Tokyo 112-8610, Japan.\\}
\footnotetext{$\mbox{}^*$E-mail address: 
pijush@degway.phys.ocha.ac.jp}  

\maketitle
\begin{abstract} 

We show that the supersymmetric rational Calogero-Moser-Sutherland (CMS) model
of $A_{N+1}$-type is equivalent to  a set of free super-oscillators,
through a similarity transformation.
We prescribe methods to construct the complete eigen-spectrum and the
associated eigen-functions, both in supersymmetry-preserving as well as
supersymmetry-breaking phases, from the free super-oscillator basis.
Further we show that a wide class of super-Hamiltonians
realizing dynamical $OSp(2|2)$ supersymmetry, which also includes all types
of rational super-CMS as a small subset, are equivalent to free
super-oscillators. We study $BC_{N+1}$-type super-CMS model in some detail
to understand the subtleties involved in this method.
\end{abstract}
\narrowtext

\newpage

\section{Introduction}

The rational CMS Hamiltonian is described by $N$
particles interacting with each other through an inverse square interaction and
all particles are subjected to a common confining harmonic force. This model
is exactly solvable and the eigen-values, including the degeneracy at each
level, are exactly identical to the spectrum of $N$ free oscillators, except
for a constant shift in the ground state energy\cite{cs,cs1,pr,bh,poly}. There
were enough indications in the literature in different context that a very close
connection between the CMS and the free oscillators model might exists. In
fact, it has been shown recently that the rational CMS
Hamiltonian is equivalent to that of free oscillators through a similarity
transformation\cite{pani,sogo}, confirming all previous speculations. This
equivalence has
enriched our understanding of the model and also became a very useful tool
for studying different aspects of CMS, like eigen functions, integrability, and
symmetry algebra, in a new way.

The supersymmetric version of the rational CMS system has also been studied in
the literature in different context\cite{fm,turb,me,sasa,in,nw,other}. The zero
fermion sector of the supersymmetric CMS describes the usual CMS model,
while the $N$ fermion sector describes the CMS model at a shifted value of
the coupling constant \cite{me}. Such relation between the zero and the $N$
fermion sector
of the model is due to `shape invariance' of the Hamiltonian, which is a very
popular and useful concept in studying quantum mechanics with one degree of
freedom \cite{cks}. For other sectors with fermion numbers ranging from one
to $N-1$,
however, no such trivial identifications with the usual CMS can be made.
Hamiltonian in these sectors are in fact related to CMS with internal degrees
of freedom\cite{bh,in}.

The supersymmetric rational CMS model (SRCMSM) of $A_{N+1}$-type is exactly
solvable in both
supersymmetry-preserving and supersymmetry-breaking phases\cite{fm,bh,turb}. 
The spectrum in the supersymmetry-preserving phase is again identical to that
of the free super-oscillators\cite{fm}. 
It might be recalled at this point that the supersymmetry is always preserved
in the super-oscillator model, once the convention for choosing the ground
state in either zero or $N$ fermion sector has been made.
Thus, the spectrum
of SRCMSM in the supersymmetry-breaking phase, has no counter-part in the
super-oscillator model. However, it has some similarity with the spectrum of
the super-oscillator model modulo a constant shift in the ground state
energy\cite{fm}. It is intriguing at this point to ask, whether or not
the SRCMSM, at least in the supersymmetry-preserving phase, can be shown
to be equivalent to free super-oscillators through a similarity transformation,
much akin to its non-supersymmetric version.

The purpose of this paper is to show that the SRCMSM of $A_{N+1}$-type is
indeed equivalent to free super-oscillators through a similarity transformation.
This equivalence is valid only in supersymmetry-preserving phase. This explains
the identicalness of the spectrum of SRCMSM and free super-oscillators. 
The eigen functions of these two models are of-course different from each
other and we outline a method to construct eigen functions of the
SRCMSM from permutationally invariant super-oscillator basis functions.
In case, one chooses a basis function which is not symmetric under the combined
exchange of bosonic and fermionic coordinates, the corresponding eigenfunction
of the SRCMSM is not normalizable. This is due to highly correlated nature
of the many-body inverse-square interaction and this has also been observed
in the usual CMS model\cite{pani}.

We also prescribe on constructing eigen-spectrum of the SRCMSM
in the supersymmetry-breaking phase, from the known
super-oscillator basis by making use of a duality property
of the model\cite{fm}. In particular, we construct a new super-Hamiltonian,
which differs from the SRCMSM by the fermionic number operator and
a constant. This implies that any eigen-function of this dual model is also
a valid eigen-function of the SRCMSM. Of course, the corresponding energy
eigen-values are different from each other. We show
through a similarity transformation that this dual Hamiltonian is again
equivalent to a free super-oscillator Hamiltonian. It turns out that the
eigen-spectrum of the SRCMSM obtained from this super-oscillator model
via the dual Hamiltonian indeed correctly describes the supersymmetry-breaking
phase of the model.

The symmetry algebra 
of the super-oscillator model is well understood in terms of a set of bosonic
and fermionic operators. We define a set of such operators for the
SRCMSM, which are obtained from the corresponding operators in
the super-oscillator model through the inverse similarity transformation. 
This enables us to study the symmetry algebra of SRCMSM in a simple way,
leading to the construction of the complete eigen-spectrum algebraically.

As a generalization of these results, we show that a wide class of models
whose bosonic many-body potential is a
homogeneous function of degree $-2$ and all the particles are restricted
to move on a line by a common confining harmonic force, are equivalent
to the super-oscillator model through a similarity transformation.
These Hamiltonians are characterized by a dynamical $OSp(2|2)$ supersymmetry.
The SRCMSM associated with different root-structures of the Lie-algebra
appear as a special small subset of this class. Though the equivalence is
valid at the operator level for the general inverse-square potential, 
one must show that the complete set of eigen-functions as well as eigen-values
of such models are indeed obtained from the super-oscillator model. The
equivalence relation at the operator level acts as a necessary condition,
while the construction of the complete eigen-spectrum and associated
wave-functions from the super-oscillator basis is sufficient to claim
such relation between these two models. We show that both the necessary
and the sufficient conditions are certainly satisfied by
the SRCMSM of $A_{N+1}$ and $BC_{N+1}$ types. However, it appears
that all other cases have to be treated individually. 

We organize the paper in the following way. We first give an overview 
of the supersymmetric quantum mechanics with many degrees of freedom
in the next section. We mostly review the known results in a way
which will become useful for our subsequent discussions. In Sec. III,
we consider the $A_{N+1}$-type SRCMSM and show its equivalence
to free super oscillator model. We first show the equivalence for the
supersymmetry-preserving phase in Sec. III.A and outline a method to
construct the eigen-spectrum from the known super-oscillator basis.
Similar study for the supersymmetry-breaking phase has been discussed
in Sec. III.B, using a duality property of the model. We generalize these
results to SRCMSM associated with other root-structures of the Lie
algebra in Sec. IV. Finally, in Sec. V, we summarize and
discuss the implications of these results. We show how the dynamical
$OSp(2|2)$ supersymmetry is realized by these systems in Appendix A.

\section{Supersymmetric Quantum Mechanics with many degrees of freedom: brief review }

The supercharge $Q$ and its conjugate $Q^{\dagger}$ are defined as,
\be
Q=\sum_{i=1}^N \psi_i^{\dagger} \ a_i, \ \ \ \ 
Q^\dagger = \sum_{i=1}^N \psi_i \  a_i^{\dagger},
\label{eq0}
\ee
\noindent where the fermionic variables $\psi_i$'s satisfy the Clifford algebra,
\be
\{\psi_i,\psi_j\}=0=\{\psi_i^{\dagger},\psi_j^{\dagger}\}, \ \
\{\psi_i, \psi_j^{\dagger}\}=\delta_{ij}, \ \ i, j=1, 2, \dots, N.
\label{eq1}
\ee
\noindent The operators $a_i(a_i^{\dagger})$'s are analogous to bosonic 
annihilation ( creation ) operators. They 
are defined in terms of the momentum operators
$p_i=-i \frac{\partial}{\partial x_i}$ and the superpotential
$W(x_1, x_2, \dots, x_N)$ as,
\be
a_i= p_i - i W_i, \ \ a_i^{\dagger}= p_i + i W_i, \ \ 
W_i = \frac{\partial W}{\partial x_i},
\label{eq2}
\ee
\noindent and satisfy the following commutation relations among themselves,
\be
[a_i,a_j]=0=[a_i^{\dagger}, a_j^{\dagger}], \ [a_i, a_j^{\dagger}]=
[a_j, a_i^{\dagger}]= 2 W_{ij},
\ \ W_{ij}= \frac{\partial^2 W}{\partial x_i \partial x_j}.
\label{eq3}
\ee
\noindent Note that, by construction,  $W_i$'s satisfy the so called
`zero-curvature
condition' $\partial_i W_j = \partial_j W_i$. Also, for translationally
invariant superpotential, these $W_i$'s satisfy the `sum to zero' condition,
$\sum_i W_i=0$. These two properties are useful ingredients in studying the
usual CMS model.

The supersymmetric Hamiltonian is defined in terms of the supercharges as,
\bea
H & = & \frac{1}{2} \{Q,Q^{\dagger}\}\nonumber \\
&= & \frac{1}{4} \sum_i \{a_i,a_i^{\dagger}\} + \frac{1}{4}
\sum_{i,j} [a_i,a_j^{\dagger}] [\psi_i^\dagger, \psi_j].
\label{eq4}
\eea
\noindent The Hamiltonian commutes with both $Q$ and $Q^{\dagger}$. The ground
state of $H$ is annihilated by both $Q$ and $Q^{\dagger}$.
Thus, the ground states are given by,
\be
\phi_0 = e^{-W} |0>, \ \ \phi_N = e^{W} |\bar{0}>,
\label{eq5}
\ee
\noindent where the fermionic vacuum $|0>$ and its conjugate
$|\bar{0}>$ in the $2^N$ dimensional fermionic Fock space are defined as,
\be
\psi_i |0> =0 , \ \ \ \psi_i^{\dagger} |\bar{0}>=0.
\label{eq6}
\ee
\noindent The first equation of (\ref{eq6}) defines the zero-fermion sector,
while the second one defines the $N$ fermion sector. In case, either $\phi_0$
or $\phi_N$ is normalizable, the
supersymmetry is preserved with zero ground state energy. On the other
hand, the supersymmetry is broken if neither $\phi_0$ nor $\phi_N$ is
normalizable. The ground state energy in this case is positive-definite.

\section{Equivalence : Rational CMS of $A_{N+1}$-type and free oscillators}

The superpotential for the $A_{N+1}$-type SRCMSM is given by,
\be
W=- \lambda ln \prod_{i<j} x_{ij} + \frac{1}{2} \sum_i x_i^2, \ \
x_{ij}=x_i-x_j.
\label{eq6.1}
\ee
\noindent The first term produces the many-body inverse square interaction,
while the second term generates the term responsible for harmonic confinement.
The Hamiltonian (\ref{eq4}), with the above choice of $W$, has the following
form,
\bea
H & = & - \frac{1}{2} \sum_i \frac{\partial^2}{\partial x_i^2} +
\frac{1}{2} \lambda (\lambda-1) \sum_{i \neq j} x_{ij}^{-2} +
\frac{1}{2} \sum_i x_i^2 - \frac{1}{2}  N \left (  1 +
\lambda (N-1) \right )\nonumber \\
& + &\sum_i \psi_i^{\dagger} \psi_i  +
\lambda \sum_{i \neq j} x_{ij}^{-2} \left ( \psi_i^{\dagger} \psi_i -
\psi_i^{\dagger} {\psi_j} \right ).
\label{eq6.2}
\eea
\noindent The Hamiltonian $H$ is permutationally invariant under the combined
exchange of bosonic and fermionic coordinates. Observe that the zero-fermion
sector of (\ref{eq6.2}) describes the usual CMS, apart from a constant equal
to its ground state energy. The ground state of SRCMSM has the well-known form,
\bea
\Phi &  = & e^{-W} |0>\nonumber \\
&=& \prod_{i<j} x_{ij}^{\lambda} e^{-\frac{1}{2} \sum_i x_i^2} |0>.
\label{eq6.3}
\eea
\noindent Note that $\Phi$ is normalizable
for $\lambda > -\frac{1}{2}$. However, a stronger criteria that each momentum
operator $p_i$ is self-adjoint for the wave-functions of the form $\Phi$
requires $ \lambda > 0$. The supersymmetry is preserved
for $\lambda > 0$, while it is broken for $\lambda < 0$\cite{fm}.

\subsection{Supersymmetry-preserving phase}

Now we would like to show that the Hamiltonian (\ref{eq6.2}) is equivalent
to the free super-oscillators model through a similarity transformation. In
order to do so, let us first consider the following transformation,
\bea
H_1 & = & e^{W} H e^{-W}\nonumber \\
& = & \sum_i \left ( x_i \frac{\partial}{\partial x_i} +
\psi_i^{\dagger} \psi_i \right ) - S ,\\
S & = & \frac{1}{2} \sum_i \frac{\partial^2}{\partial x_i^2} + 
\lambda \sum_{i \neq j} x_{ij}^{-1} \frac{\partial}{\partial x_i} -
\lambda \sum_{i \neq j} x_{ij}^{-2} \left ( \psi_i^{\dagger} \psi_i -
\psi_i^{\dagger} \psi_j \right ).
\label{eq6.4}
\eea
\noindent The total fermion number operator $N_f=\sum_i \psi_i^{\dagger}\psi_i$
commutes with the Hamiltonian $H$. The fermionic part of $H_1$ is identical
to that of $H$. Thus, $N_f$ commutes with $H_1$ and hence, also with $S$.
Making use of the following identities,
\be
\left [ \sum_i x_i \frac{\partial}{\partial x_i}, S \right ] = - 2 S, \ \ \
\left [ \sum_i \left ( x_i \frac{\partial}{\partial x_i} +
\psi_i^{\dagger} \psi_i \right ) , S \right ] = - 2 S, 
\label{eq6.5}
\ee
\noindent we find,
\bea
&& \left [ H_1, e^{-\frac{S}{2}} \right ] = S e^{-\frac{S}{2}} \\
&& H_2 = e^{\frac{S}{2}} H_1 e^{-\frac{S}{2}}\nonumber \\
&& \ \ \ \ = \sum_i \left ( x_i \frac{\partial}{\partial x_i} +
\psi_i^{\dagger} \psi_i \right ).
\label{eq6.6}
\eea
\noindent The transformed Hamiltonian $H_2$ is nothing but the supersymmetric
generalization of the Euler operator. The connection of $H$ with the free
super-oscillators is apparent from the expression of $H_2$. In particular,
we get the familiar supersymmetric $N$ particle free oscillators model in
the following way,
\bea
H_{sho} & = & e^{-\frac{1}{2} \sum_i x_i^2 } e^{-
\frac{1}{4} \sum_i \frac{\partial^2}{\partial x_i^2}} H_2
e^{\frac{1}{4} \sum_i \frac{\partial^2}{\partial x_i^2}} 
e^{\frac{1}{2} \sum_i x_i^2 }\nonumber \\
& = &  \frac{1}{2} \sum_i \left ( - \frac{\partial^2}{\partial x_i^2}
+ x_i^2 \right ) + \sum_i \psi_i^{\dagger} \psi_i - 
\frac{N}{2}.
\label{eq6.7}
\eea
\noindent This shows the equivalence between SRCMSM and
the free super-oscillators.

\subsubsection{Construction of eigen-functions}

The eigen-spectrum of (\ref{eq6.2}) can be constructed either from
(\ref{eq6.6}) or (\ref{eq6.7}). We prefer to work with Eq. (\ref{eq6.6}).
If $P_{n,k}$ is an eigen-function of (\ref{eq6.6}) with the eigen-value
$E_{n,k}$, then, $H$ has the same eigen-value $E_{n,k}$ with the
eigen-function given by,
\be
\chi= e^{-W} e^{-\frac{S}{2}} P_{n,k} \ |0>.
\label{eq6.8}
\ee
\noindent We have to choose $P_{n,k}$ to be a permutationally symmetric
 polynomial of
$x_i$ and $\psi_i$, under the
combined exchange of the bosonic and fermionic coordinates. Otherwise, the
action of $S$ on $P_{n,k}$ produces non-vanishing singular terms, thereby,
making $\chi$ non-normalizable. It is worth recalling at this point that similar
constraint on $P_{n,k}$ has been noticed also for the usual CMS case,
reflecting the highly correlated nature of these systems. 
The highly correlated nature of this model is also present in the
supersymmetric version.

There are many choices for the polynomial $P_{n,k}$. Let us choose
the following form of $P_{n,k}$,
\be
P_{n,k}= r^{2 n} \sum_i x_i^{k-1} \psi_i^{\dagger}, \ \ \
r^2=\sum_i x_i^2,
\label{eq6.9}
\ee
\noindent as the $N_f=1$ solution of $H_2$ with $E_{n,k}= 2 n + k$. The quantum
numbers $n$ and $k-1$ are nonnegative integers. It
can be checked easily that the action of $S^m$ on $P_{n,k}$ does not produce
any singularity for positive $m$. Let us first consider the action of $S$ on
$P_{n,k}$,
\bea
S P_{n,k} & = & b_1 r^{2 (n-1)} \sum_i x_i^{k-1} \psi_i^{\dagger}
+ b_2 r^{2 n} \sum_i x_i^{k-3} \psi_i^{\dagger}
+ \lambda r^{2n} \sum_{i \neq j} \sum_{l=0}^{k-3} (k-l-2) x_i^{k-l-3}
x_j^{l} \psi_i^{\dagger},\nonumber \\ 
b_1 & = & n \left [ N + 2 \lambda N (N-1) + 2 ( n + k - 2) \right ], \ \
b_2 = \frac{1}{2} (k-1) (k-2).
\label{eq6.9.1}
\eea
\noindent The first two terms on the right hand side of the first equation in
(\ref{eq6.9.1}) has the same form as that of $P_{n,k}$, except for powers
of $r$ and $x_i$. Thus, the contribution of these two terms to $S^2 P_{n,k}$
can not contain a singular term. The third term has a different form than
$P_{n,k}$. A term like this, which has a general form,
\be
\eta  = \sum_{i_{1} \neq i_{2} \neq \dots, \neq i_N } x_{i_1}^{k_1}
x_{i_2}^{k_2} \dots x_{i_N}^{k_N} \ \psi_{i_1}^{\dagger},
\label{eq6.9.2}
\ee
\noindent keeps on appearing on each successive operation of $S$ on the
left hand side of the first equation of (\ref{eq6.9.1}). The 
integers $k_i$'s are determined in terms of $k$. However, we keep them as
arbitrary nonnegative integers in (\ref{eq6.9.2}). The first term of $S$,
a generalized Laplacian operator $\bigtriangledown =
\sum_i \frac{\partial^2}{\partial x_i^2}$ ,
acting on $\eta$ can not produce any singularity. Further, we have the
following identity,
\bea
S^{\prime} \eta & = & \left ( S - 
\frac{1}{2} \sum_i \frac{\partial^2}{\partial x_i^2} \right ) \eta \nonumber \\
& = & \lambda \sum_{i_{1} \neq  i_{2} \neq \dots \neq i_N}
\sum_{j (\neq i_p)} \sum_{p=1}^N
\sum_{l=0}^{k_p-2} \ \beta_{k_p,l} \ x_{i_1}^{k_1} x_{i_2}^{k_2} \dots
x_{i_p}^{k_p-l-2} x_j^{l} \dots x_{i_N}^{k_N} \
\psi_{i_1}^{\dagger},\nonumber \\
&& \beta_{k_1,l} = k_1 - l - 1, \ \ \beta_{k_p,l} = \frac{k_p}{2} \ for \ 
p \geq 2.
\label{eq6.9.3}
\eea
\noindent Note that $S^{\prime} \eta$ has the same form as that of $\eta$,
once the summation over the indices $j, p$ and $l$ has been performed.
This proves that $S^m P_{n,k}$ can not contain a singular term, instead
terminates as a finite degree polynomial. Thus,
the well-behaved eigen-functions $\chi$ of $H$ can be constructed
from the super-oscillator basis $P_{n,k}$. It may be worth mentioning here
that the exact solution for $N_f=1$ and certain small values of $k$, obtained
in \cite{fm}, can be reproduced in a systematic way from Eqs. (\ref{eq6.8})
and (\ref{eq6.9}).

Similar results for other values of $N_f$ can also be obtained. For example,
one may choose $P_{n,k}$ for an arbitrary $N_f$ as,
\be
P_{n,k} = \frac{1}{N_f!} r^{2 n} 
\sum_{i_1, i_2, \dots, i_{N_f}} f_{i_{1} i_{2} \dots i_{N_f}}
(x_1, x_2, \dots, x_N) \  \psi_{i_1}^{\dagger}
\psi_{i_2}^{\dagger} \dots \psi_{i_{N_f}}^{\dagger},
\label{eq6.10.1}
\ee
\noindent where $ f_{i_{1} i_{2} \dots i_{N_f}}$ is anti-symmetric under
the exchange of any two indices and is a homogeneous function of degree
$k-N_f$. The anti-symmetric nature of $f$ ensures that $P_{n,k}$ is
permutationally invariant under the combined exchange of bosonic and
fermionic coordinates. Though we do not present here results concerning
normalizability of eigen-functions $\chi$ constructed from (\ref{eq6.10.1})
for arbitrary $N_f$,
it is expected that the certain specific choices of $f$ would indeed produce
well-behaved and physically accepted $\chi$. This is because of the result
\cite{turb} that the
eigen-value equation of $H_1$ has permutationally symmetric polynomials
in $x_i$ and $\psi_i$ as the solution.
This implies that the solution for the eigen equation of $S$ are also
permutationally symmetric polynomials. Thus, the action of $S^m$ on
these permutationally symmetric polynomials for any positive $m$ are not
expected to produce singular terms. We outline a method in the next section
to construct the eigenstates in an algebraic way.

\subsubsection{Algebraic structure}

The algebraic structure of the super-oscillators can be exploited to construct
the eigenstates of $H$ in an algebraic way. Consider the following set of
operators,
\bea
&& b_i^{-} =  i p_i = \frac{\partial}{\partial x_i}, \ \ \
b_i^{+}= 2 x_i\nonumber \\
&& B_n^{-} = \sum_{i=1}^N T^{-1} b_i^{-^n} T, \ \ \ 
B_n^{+}=  \sum_{i=1}^N T^{-1} b_i^{{+}^n} T, \ \ \
T= e^{\frac{S}{2}} \ e^{W}\nonumber \\
&& F_n^{-}= T^{-1} \left ( \sum_i \psi_i b_i^{-^{n-1}} \right ) T,\ \ \
F_n^{+}=T^{-1} \left ( \sum_i \psi_i^{\dagger}
b_i^{{+}^{n-1}} \right ) T,\nonumber \\
&& q_n^{-}=T^{-1} \left ( \sum_i \psi_i^{\dagger} b_i^{-^n} \right ) T, \ \ \
q_n^{+}= T^{-1} \left ( \sum_i \psi_i b_i^{{+}^n} \right ) T.
\label{eq6.10}
\eea
\noindent Note that we are using a particular form of $b_i^-$ and $b_i^+$,
such that $[b_i^-, b_j^+] = 2 \delta_{ij}$. This choice has been made to
make one to one correspondence between the usual annihilation (creation)
operator of the harmonic oscillator and the $b_i^-(b_i^+)$. In particular,
it can be checked easily,
\be
- i b_i^{-} = t^{-1} a_h^{-} t, \ \
i b_i^{+} = t^{-1} a_h^{+} t, \ \
a_h^{\pm} = p_i \pm i x_i, \ \
t= e^{- \frac{1}{2} \sum_i x_i^2} e^{ - \frac{1}{4}
\sum_i \frac{\partial^2}{\partial x_i^2}}.
\ee
\noindent The operators in (\ref{eq6.10}) satisfy the following algebra among
themselves.
\bea
&& \{F_m^{+}, F_n^{+}\}=0, \ \ [B_m^{+},F_n^{+}]=0, \ \
[B_m^{+}, B_n^{+}]=0,\nonumber \\
&& \{q_1^{-}, F_n^{+}\}=0, \ \ 
\{q_1^{+}, F_n^{+}\}=B_n^{+}, \ \ 
[H, F_n^{+}]=n F_n^{+},\nonumber \\ \ \ 
&& [q_1^{-},B_n^{+}]= 2 n F_n^{+}, \ \ [q_1^{+}, B_n^{+}]=0, \ \
[H, B_n^{+}]= n B_n^{+}.
\label{eq6.11}
\eea
\noindent This is also the algebra of the corresponding operators of
super-oscillators. Thus, the eigen-functions can be created in a
similar way by acting different powers of $B_n^{+}$ and 
$F_n^{+}$ on the ground state. In particular\cite{fm},
\be
\chi_{n_1 \dots n_{N} \nu_1 \dots \nu_N}= \prod_{k=1}^N B_k^{+^{n_k}} 
F_k^{+^{\nu_k}} \ \Phi,
\label{eq6.11.1}
\ee
\noindent is the eigenfunction with the eigen-value $E= \sum_{k=1}^{N}
k (n_k + \nu_k)$. The bosonic quantum numbers $n_k$'s are nonnegative
integers, while the fermionic quantum numbers $\nu_k$'s are either
$0$ or $1$. Note that a set of $N$ independent super-oscillators
with the frequencies $1, 2, \dots, N$ have the same energy $E$. Thus,
the spectrum of SRCMSM is identical to that of $N$ independent
super-oscillators with the frequencies $1, 2, \dots, N$.

A particular realization of the operators $B_2^+$, $B_3^+$, $F_2^+$ and
$F_3^+$ was obtained in \cite{fm}. One can easily check that the explicit 
forms of these operators found in \cite{fm}, are indeed identical to those
obtained from (\ref{eq6.10}). This equivalence is valid modulo an overall
normalization factor. Thus, we have given a systematic way to determine
$B_n^{+}$ and $F_n^{+}$ for arbitrary $n$. It might be noted here that the
particular basis we choose for the definitions of these operators is
over-complete. However, one may always choose a basis similar to one given
in \cite{turb} to avoid the over-completeness.

\subsection{Supersymmetry-breaking phase}

Consider the following supercharges,
\be
\tilde{Q}=\sum_i \psi_i \left ( p_i -i \tilde{W}_i \right ), \ \
{\tilde{Q}}^{\dagger}=\sum_i \psi_i ^{\dagger} \left ( p_i +
i \tilde{W}_i \right ), \ \
\tilde{W} = \lambda ln \prod_{i <j} x_{ij} + \frac{1}{2} \sum_i x_i^2.
\label{eq6.13}
\ee
\noindent These supercharges can be obtained from Eqs. (\ref{eq0}) and
(\ref{eq6.1}) by making  $\lambda \rightarrow - \lambda$ and $\psi_i
\leftrightarrow \psi_i^{\dagger}$. The dual Hamiltonian
$H_{d}=\frac{1}{2} \{\tilde{Q}, {\tilde{Q}}^{\dagger} \}$ differs from
$H$ by the fermionic number operator $N_f$ and a constant. In particular,
\bea
H_d & = & - \frac{1}{2} \sum_i \frac{\partial^2}{\partial x_i^2} +
\frac{1}{2} \lambda (\lambda-1) \sum_{i \neq j} x_{ij}^{-2} +
\frac{1}{2} \sum_i x_i^2 + \frac{1}{2}  N \left (  1 +
\lambda (N-1) \right )\nonumber \\
& - &\sum_i \psi_i^{\dagger} \psi_i  +
\lambda \sum_{i \neq j} x_{ij}^{-2} \left ( \psi_i^{\dagger} \psi_i -
\psi_i^{\dagger} {\psi_j} \right ),\nonumber \\
H & = & H_d + 2 N_f - N \left ( 1 + \lambda ( N-1) \right ).
\label{eq6.14}
\eea
\noindent The ground state of $H_d$ is in the $N$ fermion sector,
\be
{\tilde{\Phi}}= e^{-\tilde{W}} \ |\bar{0}>=
\prod_{i <j} x_{ij}^{-\lambda} e^{-\frac{1}{2} \sum_i x_i^2} \ |\bar{0}>,
\label{eq6.15}
\ee
\noindent which is normalizable for $ \lambda < \frac{1}{2}$. A stronger
criteria that each momentum operator $p_i$ is self-adjoint for wave-functions
of the form $\tilde{\Phi}$ determines $\lambda < 0$. The supersymmetric
phase of $H_d$ is described by $\lambda < 0$. The wave-function
$\tilde{\Phi}$ is also an eigen-state of $H$ with positive energy.
This is, in fact, the ground state of $H$ in the supersymmetry-breaking
phase\cite{fm}. The complete spectrum of $H$ in this phase can be obtained
from $H_d$ by making use of the second equation of (\ref{eq6.14}).

We get the super-oscillator Hamiltonian under the following transformations,
\bea
{\tilde{H}}_2 & = & e^{\frac{S}{2}} e^{\tilde{W}} H_d e^{-
\tilde{W}} e^{-\frac{S}{2}}\nonumber \\
&=& \sum_i \left ( x_i \frac{\partial}{\partial x_i} - 
\psi_i^{\dagger} \psi_i \right ) + N\nonumber \\
{\tilde{H}}_{sho} & = & e^{-\frac{1}{2} \sum_i x_i^2 } e^{-
\frac{1}{4} \sum_i \frac{\partial^2}{\partial x_i^2}} {\tilde{H}}_2
e^{\frac{1}{4} \sum_i \frac{\partial^2}{\partial x_i^2}} 
e^{\frac{1}{2} \sum_i x_i^2 }\nonumber \\
& = &  \frac{1}{2} \sum_i \left ( - \frac{\partial^2}{\partial x_i^2}
+ x_i^2 \right ) - \sum_i \psi_i^{\dagger} \psi_i + \frac{N}{2}.
\label{eq6.16}
\eea
\noindent Note the difference between $H_{sho}$ and ${\tilde{H}}_{sho}$.
The ground state is in the $N_f=0$ sector for the former case, while it
is in the $N_f=N$ sector for the latter one. This is expected also, since
the original many-body Hamiltonians $H$ and $H_d$ have ground states 
in the $N_f=0$ and $N_f=N$, respectively.

We use the first equation of (\ref{eq6.16}) to construct eigen-spectrum
of $H$. The eigen-function is given by,
\be
{\hat{\Phi}} = e^{-\tilde{W}} e^{- \frac{S}{2}} {\tilde{P}}_{n,k} |\bar{0}>,
\label{eq6.17}
\ee
\noindent where ${\tilde{P}}_{n,k}$ is a permutationally invariant polynomial
under the combined exchange of $x_i$ and $\psi_i$. We may choose
${\tilde{P}}_{n,k}$ to have the same form as $P_{n,k}$, except for the 
replacement $\psi_i^{\dagger} \rightarrow \psi_i$. Following the discussions
on the supersymmetry-preserving phase in Sec. III.A.1, it can be checked
easily that this choice of ${\tilde{P}}_{n,k}$ results in well-behaved,
normalizable eigenfunction for $H$.

The complete eigenstates can also be constructed with the help of bosonic
creation operator $\hat{B}_n^+$ and the fermionic creation operator
$\hat{F}_n^+$. We define,
\be
\hat{B}_n^+ = \sum_i \hat{T}^{-1} b_i^{+^n} \hat{T}, \ \
\hat{F}_n^{+} = \hat{T}^{-1} \left (\sum_i
\psi_i b_i^{+^{n-1}} \right ) \hat{T},\ \
\hat{q}_n^{+} = \hat{T}^{-1} \left ( \sum_i \psi_i^{\dagger}
b_i^{+^n} \right ) \hat{T},
\label{eq6.17.1}
\ee
\noindent with $\hat{T}= e^{\frac{S}{2}} e^{\hat{W}}$. The eigenstates
are,
\be
\hat{\Phi}_{n_1, \dots, n_N, \nu_1, \dots, n_N} =
\prod_{k=1}^N \hat{B}_k^{+^{n_k}} 
\hat{F}_k^{+^{\nu_k}} \tilde{\Phi} ,
\ee
\noindent with the eigen-values, $E=N ( 1 - \lambda (N-1) ) +
\sum_{k=1}^N (k n_k + (k-2) n_k)$. The bosonic quantum numbers $n_k$'s
are non-negative integers, while the fermionic quantum numbers $\nu_k$'s
are either $0$ or $1$.

\section{Generalization}

We have constructed a similarity transformation which shows the equivalence
between the SRCMSM and free super-oscillators. The particular SRCMSM
we considered is associated with the $A_{N+1}$ type root-structure of the Lie
algebra. SRCMSM associated with other root structures also can be shown
to be equivalent to free super-oscillators. Instead of considering each model
separately, we prove below a general result, which is applicable to all
types of SRCMSM and also to a new class of rational models considered
in \cite{kha} having nearest-neighbor and next-nearest-neighbor interactions.
In particular, we consider a super-Hamiltonian ${\cal{H}}$ whose
bosonic many-body
potential is a homogeneous function of degree $-2$ and all the particles are
confined on the line by a common harmonic oscillator potential. It is worth
recalling at this point that all types of SRCMSM and models considered
in \cite{kha}, indeed satisfy this criteria. We construct a
similarity transformation which shows the equivalence between ${\cal{H}}$
and free super-oscillators $H_{sho}$.

Let us decompose the superpotential $W$ in terms of superpotentials for
the many-body interaction and the harmonic term as,
\be
W= - w + \frac{1}{2} \sum_i x_i^2, \ \ w= ln G(x_1, x_2, \dots, x_N),
\label{eq7}
\ee
\noindent where $G$ is a homogeneous function of any arbitrary positive
degree $d$,
\be
\sum_i x_i \frac{\partial G}{\partial x_i} =  d G.
\label{eq7.01}
\ee
\noindent This property
of $G$ ensures that each $w_i$ is a homogeneous function of degree $-1$ and
hence, the bosonic potential is always homogeneous function of
degree $-2$, apart from the harmonic term.
The Hamiltonian is given by,
\be
{\cal{H}}= \frac{1}{2} \sum_i \left [ - \frac{\partial^2}{\partial x_i^2} +
w_i^2 + w_{ii} + x_i^2 \right ] - ( d + \frac{N}{2} )
+ \sum_i \psi_i^{\dagger} \psi_i - \sum_{i,j} w_{ij} \psi_i^{\dagger} \psi_j.
\label{eq7.1}
\ee
\noindent This Hamiltonian has a dynamical $OSp(2|2)$ supersymmetry.
The full $OSp(2|2)$ algebra and the operators realizing this algebra are
given in Appendix-A. 

The bosonic sub-algebra $O(2,1) \times U(1)$ of $OSp(2|2)$ is present
for a wide class of Hamiltonians ${\cal{H}}$, due to the constraint
(\ref{eq7.01}) on the superpotential. This class of Hamiltonians having
$O(2,1) \times U(1)$ symmetry can even
be made larger by adding a term $T$ having the following properties,
\be
[N_f, T] = 0, \ \ \left [\sum_i x_i \frac{\partial}{\partial x_i},
T \right ] = - 2 T,
\ee 
\noindent to the Hamiltonian ${\cal{H}}$. However, the new Hamiltonian
${\cal{H}}^{\prime}= {\cal{H}}+T$ will not be supersymmetric anymore
for general $T$. It is worth mentioning at this point that
the presence of the symmetry algebra $O(2,1) \times U(1)$ is enough to show
the equivalence between ${\cal{H}}^{\prime}$ and free super-oscillators. The
supersymmetry of the Hamiltonian does not play any role. In
other words, our results are valid even if the $OSp(2|2)$ symmetry of
${\cal{H}}^{\prime}$ is lost, but, has only $O(2,1) \times U(1)$ symmetry. 
However, we restrict our discussions in this paper to 
$OSp(2|2)$ supersymmetric Hamiltonian ${\cal{H}}$ only.

Observe that ${\cal{H}}$ can be transformed to a new Hamiltonian
${\cal{H}}_1$ under the following similarity transformation,
\bea 
{\cal {H}}_1 & = & e^{W} {\cal{H}} e^{-W}\nonumber \\
& = & \sum_i \left ( x_i \frac{\partial}{\partial x_i} +
\psi_i^{\dagger} \psi_i \right ) - \hat{S}\\
\hat{S} & = & \sum_i \left ( \frac{1}{2} \frac{\partial^2}{\partial x_i^2} +
w_i \frac{\partial}{\partial x_i} \right ) +
\sum_i w_{ii} \psi_i^{\dagger}\psi_i
+ \sum_{i \neq j} w_{ij} \psi_i^{\dagger} \psi_j .
\label{eq8}
\eea
\noindent The total fermion number $N_f=\sum_i \psi_i^{\dagger} \psi_i$
commutes with $\hat{S}$, $[N_f,\hat{S}]=0$. The commutation relation between
the Euler operator $E=\sum_i x_i \frac{\partial}{\partial x_i}$ and $\hat{S}$
is given by,
\be
[E, \hat{S}] = - 2 \hat{S}, \ \
[H_2, \hat{S}]=[E+N_f,\hat{S}]=- 2 \hat{S}.
\label{eq9}
\ee
\noindent The homogeneity property (\ref{eq7.01}) of $G$ has been used in
deriving the above equations.
Now it is easy to show that ${\cal{H}}_1$ is transformed to $H_2$ under
the following transformation,
\be
H_2= e^{\frac{1}{2} \hat{S}} {\cal{H}}_1 e^{- \frac{1}{2} \hat{S}} .
\label{eq11}
\ee
\noindent The super-oscillator Hamiltonian can be obtained from (\ref{eq11})
by using the same transformation as used in equation (\ref{eq6.7}).
 
One might wonder at this point that any super-Hamiltonian with the
superpotential $W$ described by (\ref{eq7}) and (\ref{eq7.01}) is exactly
solvable, due to its equivalence to free super oscillators through 
similarity transformations. We would like to point out that this may
not be true always, because, merely showing the equivalence of different
models is not sufficient for such conclusions. We have to make sure that
the similarity transformation, which is responsible for such equivalence,
keeps the original Hamiltonian in its own Hilbert space. Thus, as a check,
one should show that the complete spectrum and the corresponding
well-behaved, normalizable eigen-functions of ${\cal{H}}$ can
be constructed from $H_{sho}$ or $H_2$ through inverse similarity
transformation. The equation (\ref{eq11}) act as a necessary
condition, while the construction of the complete spectrum and associated
well-behaved eigen-functions of the original Hamiltonian from the
super-oscillator model is sufficient to claim the equivalence between
these two Hamiltonians. We discuss these points below with the example of
$BC_{N+1}$-type SRCMSM.

\subsection{$BC_{N+1}$-type SRCMSM and super-half-oscillator}

The superpotential for the $BC_{N+1}$-type SRCMSM is described by,
\be
G (\lambda, \lambda_1, \lambda_2) = \prod_{i <j} \left ( x_i^2 -
x_j^2 \right )^{\lambda} 
\prod_k x_k^{\lambda_1}  \prod_l (2 x_l)^{ \lambda_2},
\label{eq12}
\ee
\noindent where $\lambda$, $\lambda_1$ and $\lambda_2$ are arbitrary
parameters. The $D_{N+1}$-type model is described by
$\lambda_1=\lambda_2=0$, while $\lambda_1=0 (\lambda_2=0)$ describes
$C_{N+1} (B_{N+1})$-type Hamiltonian. Without loss of any generality,
we restrict our discussions to the $B_{N+1}$-type Hamiltonian only. The
Hamiltonian is given by,
\bea
H_{B_{N+1}} & = & - \frac{1}{2} \sum_i \frac{\partial^2}{\partial x_i^2} +
\frac{1}{2} \lambda (\lambda-1) \sum_{i \neq j} \left [ x_{ij}^{-2} 
+ (x_i + x_j )^{-2} \right ] + \frac{1}{2}
\lambda_1 ( \lambda_1-1) \sum_i x_i^{-2}\nonumber \\
&& + \frac{1}{2} \sum_i x_i^2
- \frac{1}{2}  N \left [  1 + 2 \lambda (N-1) + \lambda_1 \right ]
+ \sum_i \psi_i^{\dagger} \psi_i  + \lambda_1 \sum_i \psi_i^{\dagger}
\psi_i x_i^{-2}\nonumber \\
&& + \lambda \sum_{i \neq j} \left [ x_{ij}^{-2}
\left ( \psi_i^{\dagger} \psi_i - \psi_i^{\dagger} {\psi_j} \right ) + 
(x_i + x_j )^{-2} \left ( \psi_i^{\dagger} \psi_i +
\psi_i^{\dagger} {\psi_j} \right ) \right ].
\label{13}
\eea
\noindent The many-body potential is not translationally invariant like 
$A_{N+1}$-type SRCMSM. Each particle interacts with the images of
all other particles and also with itself. This kind of Hamiltonians are
suitable for describing systems with boundaries. We choose to work
in the $0 < x_1 < x_2 < \dots < x_N$ sector of the phase space. Solutions
in other sectors can be obtained by using the fact that the
Hamiltonian is permutationally invariant under the combined
exchange of $x_i$ and $\psi_i$. The Hamiltonian also has a very interesting
discrete symmetry. It is invariant under any pair
$ ( x_i , \psi_i ) \rightarrow ( - x_i, - \psi_i ) $.
This reflection symmetry has a consequence on the spectrum.

The ground-state of (\ref{13}) in the supersymmetric phase is given by,
\be
\Phi= \prod_{i <j} \left ( x_i^2 - x_j^2 \right )^{\lambda} 
\prod_k x_k^{\lambda_1}  e^{- \frac{1}{2} \sum_i x_i^2},
\ee
\noindent with $\lambda, \lambda_1 > 0$. We would like to emphasize here
that $\Phi$ is normalizable for $\lambda, \lambda_1 > -\frac{1}{2}$.
However, a stronger criteria that each momentum operator $p_i$ is
self-adjoint for the wave-function of the form $\Phi$ has been imposed.
This requires $\lambda$ and $\lambda_1$ to be positive definite.
The supersymmetry-breaking phase of the $BC_{N+1}$-type model has a richer
structure than the $A_{N+1}$-type model. In the parameter space of $\lambda$
and $\lambda_1$, there are three regions for which the supersymmetry is broken.
They are, (i) $\lambda < 0$, $\lambda_1 < 0$, (ii)
$\lambda < 0$, $\lambda_1 > 0$ and (iii) $\lambda > 0$, $\lambda_1 < 0$.
We first discuss the spectrum in the supersymmetric phase in the next section.
The spectrum in the supersymmetry-breaking phase will be discussed
subsequently.

\subsubsection{Supersymmetric phase}

The complete spectrum of $H_{B_{N+1}}$ is described by a subset of the
spectrum of super-oscillators,
\be
E_{B_{N+1}}= 2 ( n + k + N_f ), \ \  E_{sho}= 2 n + k + N_f.
\label{eq14}
\ee
\noindent At a first thought, this observation might
lead to a wrong conclusion regarding the validity of the similarity
transformation for $B_{N+1}$-type SRCMSM. This apparent contradiction
is removed once the discrete reflection symmetry of the $H_{B_{N+1}}$ 
is imposed on the eigen-functions of $H_2$. In particular, we have to choose,
\be
\hat{P}_{n,k} = \frac{1}{N_f!} r^{2 n} 
\sum_{i_1, i_2, \dots, i_{N_f}} f_{i_{1} i_{2} \dots i_{N_f}}
(x_1, x_2, \dots, x_N) \  (x_{i_1} \psi_{i_1}^{\dagger})
(x_{i_2} \psi_{i_2}^{\dagger}) \dots (x_{i_{N_f}} \psi_{i_{N_f}}^{\dagger}),
\label{eq15}
\ee
\noindent where $f$ is anti-symmetric under the exchange of any two indices
and a homogeneous function of degree $2 k$. Note that $\hat{P}_{n,k}$ is
invariant under, 
(a) $(x_i, \psi_i^{\dagger}) \leftrightarrow (x_j, \psi_j^{\dagger})$ and
(b) $ (x_i, \psi_i^{\dagger}) \rightarrow (- x_i, - \psi_i^{\dagger})$.
With this choice of $\hat{P}_{n,k}$, the eigen-value $E_{sho}^{\prime}$
of $H_2$ is identical with $E_{B_{N+1}}$, $E_{B_{N+1}}= E_{sho}^{\prime}=
2 (n+k+N_f)$. Also, the action of ${\hat{S}}^m$ on $\hat{P}_{n,k}$ does
not produce any singularity for positive $m$. Thus, $H_{B_{N+1}}$ is
equivalent to a set of free super-half-oscillators. An explanation on the
use of the term `super-half-oscillator' is in order. Note that both
$E_{B_{N+1}}$
and $E_{sho}^{\prime}$ are always even for any integer $n$, $k$ and $N_f$.
On the other hand, there is no such restriction on $E_{sho}$. It can be
both even and odd. Thus, $E_{sho}^{\prime}$ or $E_{B{N+1}}$ describes only
half of the
spectrum described by $E_{sho}$.  This is because the
eigen-value $E_{sho}$ is for $N$ super-oscillators defined on the full-line.
On the contrary, the super-Hamiltonian $H_{B_{N+1}}$ is defined only on
the positive half-line and, hence, the $E_{B_{N+1}}$ or $E_{sho}^{\prime}$
corresponds to the
eigen-value of a set of free super-oscillators on the half-line. Thus, in
analogy with the similar problem for a single particle oscillator Hamiltonian,
we use the term `super-half-oscillator'.

The eigen-spectrum also can be constructed in an algebraic way. We define
the creation and annihilation operators as,
\be
{\cal{B}}_{n}^+ = {\cal{T}}^{-1} \sum_i b_i^{+^{2 n}} {\cal{T}}, \ \
{\cal{F}}_{n}^+ = {\cal{T}}^{-1} \sum_i \psi_i^{\dagger} 
b_i^{+^{2 n-1}} {\cal{T}}, \ \
{\cal{T}} = e^{\frac{\hat{S}}{2}} e^W .
\label{eqq}
\ee
\noindent Note that these operators are invariant under (a) and (b). Thus,
the eigen-functions obtained by operating these operators on the ground-state
also are invariant under (a) and (b). The eigen-states are obtained as,
\be
{\cal{\chi}}_{n_1 \dots n_{N} \nu_1 \dots \nu_{N}}=
\prod_{k=1}^N {\cal{B}}_k^{+^{n_k}} 
{\cal{F}}_k^{+^{\nu_k}} \ \Phi,
\label{eq66}
\ee
\noindent with the energy ${\cal{E}}= \sum_{k=1}^{N} 2 k (n_k + \nu_k)$.
The bosonic quantum numbers are non-negative integers, while the fermionic
quantum numbers are $0$ or $1$. Note that the energy ${\cal{E}}$ can be
interpreted as that of $N$ independent super-half-oscillators
with the frequencies $1, 2 \dots, N$. 
 
\subsubsection{Supersymmetry-breaking phase}

The eigen-spectrum of
the Hamiltonian in the region (i) can be obtained in a similar way as
described in Sec. III.B. In particular, we construct a dual Hamiltonian
$H_{B_{N+1}}^d$ from $H_{B_{N+1}}$ by the transformations,
$\psi_i \leftrightarrow \psi_i^{\dagger}$, $\lambda \rightarrow - \lambda$
and $\lambda_1 \rightarrow -\lambda_1$. The relation between these two
Hamiltonians is given by,
\be
H_{B_{N+1}} = H_{B_{N+1}}^d + 2 N_f - N \left [ 1 + 2 \lambda (N-1) +
\lambda_1 \right ].
\label{eq16}
\ee
\noindent Using this relation, the complete eigen-spectrum of $H_{B_{N+1}}$
in the supersymmetry-breaking phase can be obtained. In particular, the
bosonic and fermionic creation operators can be obtained from (\ref{eqq})
by replacing
$\lambda \rightarrow -\lambda$, $\lambda_1 \rightarrow -\lambda_1$
and $\psi_i \leftrightarrow \psi_i^{\dagger}$. These operators acting on
the ground state of $H_{B_{N+1}}^d$ produces the eigenstates of $H_{B_{N+1}}$
with the eigen-value, $E= N (1 - 2 \lambda (N-1) - \lambda_1 )
+ \sum_{k=1}^N 2 ( k n_k + (k-1) \nu_k )$.
This method does not work for the regions (ii) and (iii) in a straightforward
way.

The method for constructing eigen-spectrum in the regions
described by (ii) and (iii) are similar. We first study
the Hamiltonian in the region (ii). The ground state wave-function
in this region is given by,
\be
\psi ( \lambda, \lambda_1) = e^{-\theta} |\bar{0}> = 
\prod_{i<j} \left ( x_i^2 -
x_j^2 \right )^{-\lambda} \prod_k x_k^{1+\lambda_1} 
e^{- \frac{1}{2} \sum_i x_i^2} |\bar{0}>,
\label{eq17}
\ee
\noindent with the ground-state energy $E= \frac{3 N}{2} - 2 \lambda
N (N-1) $. It may be noted here that $\psi (1-\lambda, \lambda_1-1)$ is
also an exact eigenstate in the $N_f=0$ sector. However, the associated
energy eigen-value is greater than $E$ for $N \geq 3$.
We would also like to point out here that the particular form of $\psi$
is due to the shape invariance of the model, relating $N_f=0$ sector to
$N_f=N$ sector \cite{me,cks}. Now we introduce a new Hamiltonian $H_3$,
which is related to $H_{B_{N+1}}$ by the following relation,
\be
H_{B_{N+1}} = H_3 + 2 N_f - \frac{N}{2} - 2 \lambda N (N-1).
\label{eq17.19}
\ee
\noindent The above relation is similar to (\ref{eq16}). However, unlike
$H_{B_{N+1}}^d$ or $H_d$ for the $A_{N+1}$-type model, $H_3$ is not
supersymmetric. Thus, we can not use the methods of supersymmetric theory
to determine the ground-state energy of $H_3$. Instead, we find by inspection
that  $\psi(\lambda, \lambda_1)$ is the zero-energy eigenstate of $H_3$.
Now one can check easily,
\be
{\tilde{H}}_2 = e^{\frac{\hat{S}}{2}} e^{\theta} H_3 e^{-\theta} e^{-
\frac{\hat{S}}{2}}, 
\label{eq17.17}
\ee
\noindent where $\hat{S}$ can be calculated from (\ref{eq8}) for
the choice of $w$ as $w=ln G(-\lambda, \lambda_1, \lambda_2=0)$.
The bosonic and fermionic creation operators can be obtained from
(\ref{eqq}) by replacing
$\lambda \rightarrow -\lambda$, $\lambda_1 \rightarrow \lambda_1$
and $\psi_i \leftrightarrow \psi_i^{\dagger}$. These operators acting on
$\psi(\lambda, \lambda_1)$ produces the eigenstates of $H_{B_{N+1}}$
in region (ii) with the eigen-value, $E= N (\frac{3}{2} -
2 \lambda (N-1) ) + \sum_{k=1}^N 2 ( k n_k + (k-1) \nu_k )$.
Finally, we
mention that the ground-state wave-function in the region (iii) is
$\psi(-(1+\lambda), (1-\lambda_1))$ with the ground-state energy
$E= \frac{3N}{2} - 2 \lambda_1 N (N-1)$. Rest of the analysis in this
region can be done in a straightforward way.

\section{Summary and Discussions}

We have constructed a similarity transformation which maps the SRCMSM
Hamiltonian of $A_{N+1}$-type to that of a supersymmetric free harmonic
oscillators. This equivalence is valid only in the supersymmetry-preserving
phase of SRCMSM. We have outlined methods for the construction of
eigen-functions of SRCMSM from the eigen-functions of super-oscillators.
Even though there is no equivalence between SRCMSM in supersymmetry-breaking
phase and super-oscillators, we are able to construct eigen-spectrum in this
phase by using a duality property of the model. We observed that only
those eigen functions of the free super-oscillators, which are symmetric
under the combined exchange of both bosonic and fermionic coordinates,
produce normalizable wave-function for the SRCMSM. This has also been
observed in the pure bosonic case. Thus, this brings out the
highly correlated nature of these systems.

We have generalized these results to a wide class of super-Hamiltonians whose
bosonic many-body interaction is a homogeneous function of degree $-2$ and all
the particles are subjected to a common harmonic confinement. These
Hamiltonians are characterized by a dynamical $OSp(2|2)$ supersymmetry.
Though this
equivalence is certainly valid at the operator level, it turns out that
the individual super-Hamiltonians should be analyzed carefully to see if the
similarity transformation is keeping the Hamiltonian in its original Hilbert
space or not. As a check to ascertain this, one should be able to construct
the complete set of eigen-values and associated eigen functions of the original
Hamiltonian from the super-oscillator model. We discussed the $BC_{N+1}$-type
SRCMSM as an example and showed its equivalence to half of the spectrum of
super-oscillators. To the best of our knowledge, this is the first instance
in the literature where the complete spectrum and the eigenstates
of the $BC_{N+1}$-type SRCMSM has been obtained. 

The SRCMSM associated with the root structures other than $A_{N+1}$ and
$BC_{N+1}$ type have not been touched upon in this paper. We believe
that permutationally symmetric super-oscillator basis with additional
symmetry requirements coming from the specific nature of the root structure,
as in the case of $BC_{N+1}$-type model, would produce the complete spectrum
and associated well-behaved eigen-functions of these models. An universal
formulation of the method described here, valid for SRCMSM associated with all
the root structures along the line of investigations carried out
in \cite{sasa,pat}, is desirable.

The equivalence relation (\ref{eq11}) is valid for a wide class of
super-models realizing $OSp(2|2)$ supersymmetry. The CMS systems form
only a small subset. We have seen that Eq. (\ref{eq11}) act as a necessary
condition for the equivalence between the original Hamiltonian and
super-oscillators. The sufficient condition for the equivalence is to construct
the complete spectrum and associated wave-functions of the original Hamiltonian
from the super-oscillator basis. Thus, it would be interesting to construct new
exactly solvable super-models using the method described here.

\acknowledgements{I would like to thank Tetsuo Deguchi for a careful reading of
the manuscript and comments. This work is supported by a fellowship 
(P99231) of the Japan Society for the Promotion of Science.}

\appendix{\section{$OSp(2|2)$ super-algebra}

We show in this appendix that the Hamiltonian ${\cal{H}}$ in (\ref{eq7.1})
has dynamical $OSp(2|2)$ supersymmetry. We first define the following four
supercharges,
\bea
&& q= \frac{1}{2} \sum_{i=1}^N \psi_i^{\dagger} ( p_i + i w_i - i x_i), \ \
q^{\dagger}= \frac{1}{2} \sum_{i=1}^N \psi_i ( p_i - i w_i + i x_i),\nonumber \\
&& {\tilde{q}} = \frac{1}{2} \sum_{i=1}^N \psi_i \left ( p_i - i w_i -
i x_i \right ), \ \
{\tilde{q}}^{\dagger} = \frac{1}{2} \sum_{i=1}^N \psi_i^{\dagger}
\left ( p_i + i w_i + i x_i \right ).
\label{a0}
\eea
\noindent The Hamiltonian ${\cal{H}}$ is given in terms of $q$ and
$q^{\dagger}$as, ${\cal{H}} = 2 \{q, q^{\dagger}\}$. The dual Hamiltonian
${\cal{H}}^d$ can be constructed in terms of $\tilde{q}$ and
${\tilde{q}}^{\dagger}$, ${\cal{H}}^d= 2 \{{\tilde{q}},
{\tilde{q}}^{\dagger}\}$. We define the following operators,
\bea
&& h = \frac{1}{2} \left ( {\cal{H}} + {\cal{H}}^d \right ), \ \
U = \frac{1}{2}  \left ( {\cal{H}} - {\cal{H}}^d \right ),\nonumber \\
&& {\cal{B}}_2 ^- = {\cal{B}}_0 -\frac{1}{4} \sum_i x_i^2
- \frac{1}{4} \left ( N + 2 E \right ), \ \
{\cal{B}}_2 ^+ = {\cal{B}}_0 - \frac{1}{4} \sum_i x_i^2
+ \frac{1}{4} \left ( N + 2 E \right ),\nonumber \\
&& {\cal{B}}_0 = \frac{1}{4} \sum_i \left ( p_i^2 + w_i^2 + w_{ii} -
2 \sum_{j} w_{ij} \psi_i^{\dagger} \psi_j \right ).
\label{a1}
\eea
\noindent The bosonic operators ${\cal{B}}_2^{\pm}$ and $h$ satisfies
the following relations,
\be
[h, {\cal{B}}_2^{\pm}] = \pm 2 {\cal{B}}_2^{\pm}, \ \
[{\cal{B}}_2^-, {\cal{B}}_2^+] = h.
\label{a2}
\ee
\noindent The commutator relation $[E,{\cal{B}}_0]=- 2 {\cal{B}}_0$
has been used in deriving the above equations. The $U(1)$ generator $U$
commutes with ${\cal{B}}_2^{\pm}$ and $h$.

The non-vanishing anticommutators among $q$, $q^{\dagger}$, $\tilde{q}$
and $\tilde{q}^{\dagger}$ are,
\be
\{q, q^{\dagger}\} = \frac{1}{2} ( h + U ) , \ \
\{\tilde{q}, \tilde{q}^{\dagger}\} = \frac{1}{2} ( h - U ), \ \
\{q^{\dagger}, \tilde{q}^{\dagger}\} = {\cal{B}}_2^{+}, \ \
\{q, \tilde{q}\} = {\cal{B}}_2^{-} . \ \
\label{a3}
\ee
\noindent Observe that the relation,
\be
{\cal{H}} = {\cal{H}}_d + 2 U,
\label{a4}
\ee
\noindent which is useful in determining the spectrum in the
supersymmetry-breaking phase, follows easily from the first two equations
of (\ref{a3}).

The other non-vanishing commutators are,
\bea
&& [{\cal{B}}_2^+, q] = - \tilde{q}^{\dagger}, \ \
[{\cal{B}}_2^+, \tilde{q}] = - q^{\dagger}, \ \
[{\cal{B}}_2^-, \tilde{q}^{\dagger}] = q, \ \
[{\cal{B}}_2^-, q^{\dagger}] = \tilde{q},\nonumber \\
&& [h, q^{\dagger}]= q^{\dagger}, \ \
[h, q]= - q, \ \
[h, \tilde{q}]=-\tilde{q}, \ \
[h,\tilde{q}^{\dagger}] = \tilde{q}^{\dagger},\nonumber \\
&& [U, \tilde{q}] = - \tilde{q}, \ \
[U, \tilde{q}^{\dagger}] = \tilde{q}^{\dagger}, \ \
[U, q^{\dagger}] = - q^{\dagger}, \ \
[U, q] = q.
\label{a5}
\eea}

\end{document}